# Wireless Silent Speech Interface Using Multi-Channel Textile EMG Sensors Integrated into Headphones

Chenyu Tang, Josée Mallah, Dominika Kazieczko, Wentian Yi, Tharun Reddy Kandukuri, Edoardo Occhipinti, Bhaskar Mishra, Sunita Mehta, *Member IEEE*, and Luigi G. Occhipinti, *Senior Member, IEEE*

*Abstract*—**This paper presents a novel wireless silent speech interface (SSI) integrating multi-channel textile-based EMG electrodes into headphone earmuff for real-time, hands-free communication. Unlike conventional patch-based EMG systems, which require large-area electrodes on the face or neck, our approach ensures comfort, discretion, and wearability while maintaining robust silent speech decoding. The system utilizes four graphene/PEDOT:PSS-coated textile electrodes to capture speech-related neuromuscular activity, with signals processed via a compact ESP32-S3-based wireless readout module. To address the challenge of variable skin-electrode coupling, we propose a 1D SE-ResNet architecture incorporating squeeze-and-excitation (SE) blocks to dynamically adjust per-channel attention weights, enhancing robustness against motion-induced impedance variations. The proposed system achieves 96% accuracy on 10 commonly used voice-free control words, outperforming conventional single-channel and non-adaptive baselines. Experimental validation, including XAI-based attention analysis and t-SNE feature visualization, confirms the adaptive channel selection capability and effective feature extraction of the model. This work advances wearable EMG-based SSIs, demonstrating a scalable, low-power, and user-friendly platform for silent communication, assistive technologies, and human-computer interaction.**

## I. INTRODUCTION

Silent Speech Interfaces (SSIs) are an emerging category of human-machine interfaces that enable speechless communication by decoding neuromuscular or bioelectrical signals associated with speech production. SSIs hold great promise in assistive communication for individuals with speech impairments, as well as in silent command interfaces for wearable and mobile computing [1, 2, 3, 4]. Various approaches have been proposed to realize SSIs, including brain-computer interfaces (BCIs) that leverage electroencephalography (EEG) or electrocorticography (ECoG) [5, 6, 7], wearable sensors such as electromyography (EMG) and strain sensors [8, 9, 10, 11], and mobile device-based solutions that utilize acoustic and optical sensing for lip or facial movement recognition [12, 13, 14]. While BCI-based SSIs offer high precision in neural decoding, they require invasive or semi-invasive setups with high computational costs. Mobile device-based SSIs, on the other hand, are convenient but rely on optical clarity, making them unsuitable in low-light or occluded conditions. Wearable sensor-based SSIs, particularly those leveraging surface EMG (sEMG), present a promising alternative due to their balance between accuracy, portability, and real-time responsiveness.

Among wearable SSIs, two main sensing modalities have been explored: strain sensors and EMG sensors. Strain sensors, typically integrated into facial masks or wearable bands, capture minute deformations in the skin or soft tissues during articulation, making them effective for speech movement tracking [15, 16, 17]. However, their dependency on precise placement and their susceptibility to mechanical drift limit their robustness in real-world scenarios [18, 19, 20]. In contrast, EMG-based SSIs directly capture neuromuscular activity associated with speech articulation, offering a more direct and robust representation of speech-related bioelectrical signals. EMG sensors do not suffer from optical dependencies (as in mobile devices) or mechanical degradation (as in strain sensors), making them particularly suitable for low-power, wearable, and continuous-use applications.

Despite these advantages, existing EMG-based SSI implementations face significant limitations. Most current solutions rely on large-area, multi-electrode patches placed on the facial, neck, or jaw regions, covering muscles such as the masseter, submental, and laryngeal muscle groups. These flexible electrode patches are typically made from silver/silver chloride (Ag/AgCl), carbon-based composites, or conductive polymers, which, while functional, suffer from limited

This work involved human subjects. Approval of the ethical, experimental procedures and protocols was granted by Department of Engineering Ethics Committee, University of Cambridge, under the reference number 566. This work was supported by UKIERI, British Council UK (G507157_G128014) and SPARC coordinated by IIT Kharagpur India (SPARC-UKIERI/2024-2025/P3357). *(Corresponding author: Luigi G. Occhipinti)*. Chenyu Tang and Josée Mallah contributed equally to this work.

Chenyu Tang, Josée Mallah, Dominika Kazieczko, Wentian Yi, Tharun Reddy Kandukuri, and Luigi G. Occhipinti are with the Cambridge Graphene Centre, University of Cambridge, Cambridge, CB3 0FA UK (e-mail: ct631@cam.ac.uk; jm2508@cam.ac.uk; dk720@cam.ac.uk; wy278@cam.ac.uk; trk25@cam.ac.uk; lgo23@cam.ac.uk).

Edoardo Occhipinti is with the UKRI Centre for Doctoral Training in AI for Healthcare, Department of Computing, Imperial College London, London, UK (email: edoardo.occhipinti16@imperial.ac.uk).

Bhaskar Mishra is with the Shitashii Innovations Pvt Ltd, India (e-mail: m.bhaskar1304@gmail.com).

Sunita Mehta is with Chitkara University Punjab and Shitashii Innovations Pvt. Ltd (sunita.mehta@chitkara.edu.in).

breathability, poor long-term wearability, and reduced user comfort [8, 20]. Moreover, the visibility and bulkiness of such patches limit their discreet use in everyday environments [18, 21]. A more compact and user-friendly alternative is to integrate EMG electrodes into existing wearable devices such as headphones, leveraging their natural placement around the ear to capture EMG signals from nearby speech-related musculature. However, integrating multi-channel dry EMG electrodes into headphone earmuffs introduces significant challenges in electrode-skin coupling stability. Compared to conventional adhesive patches, headphone-integrated dry electrodes are inherently more susceptible to signal instability due to movement artefacts, unstable contact, and localized skin-electrode impedance changes.

To address these challenges, we propose a novel wireless silent speech interface that integrates a multi-channel textile-based EMG sensing array into headphone earmuffs, coupled with a robust adaptive decoding framework. Specifically, we leverage towel electrodes, a textile-based dry electrode technology we developed in our previous study [22, 23], and design an adaptive machine learning-based decoding algorithms (see diagram in Fig. 1). Our system consists of four towel EMG electrodes embedded into the earmuff of standard headphones, interfaced with a wireless readout module for real-time signal acquisition. The towel electrode structure, featuring microscale fibrous protrusions, enhances skin-electrode coupling during wear, mitigating contact instability. To further improve signal decoding robustness, we introduce a novel 1D SE-ResNet model, incorporating squeeze-and-excitation (SE) modules to dynamically adjust per-channel attention weights based on instantaneous contact quality. This allows the network to assign higher importance to well-coupled EMG channels while suppressing noisy or unstable ones, improving decoding robustness under real-world usage conditions. Our system achieves 96% recognition accuracy on a set of 10 commonly used silent speech commands for smart control applications, significantly outperforming traditional single-channel and non-adaptive baselines.

The proposed headphone-integrated EMG-based SSI represents a major step forward in the development of practical, discreet, and user-friendly silent speech interfaces. By combining wearable textile EMG sensors with machine learning-driven adaptive channel selection, this approach paves the way for next-generation hands-free and voice-free interaction systems in assistive communication, human-computer interaction, and wearable intelligence.

## II. METHODOLOGY

### A. Hardware of the System

The proposed SSI integrates a multi-channel textile-based EMG sensor array into headphone earmuffs, coupled with a wireless readout module for real-time signal acquisition and processing. The core sensing component of the system is a set of four textile towel-based dry EMG electrodes, developed based on our previous work. These electrodes offer highly flexible, biocompatible, and breathable sensing surfaces, optimized for stable skin-electrode coupling during long-term wear.

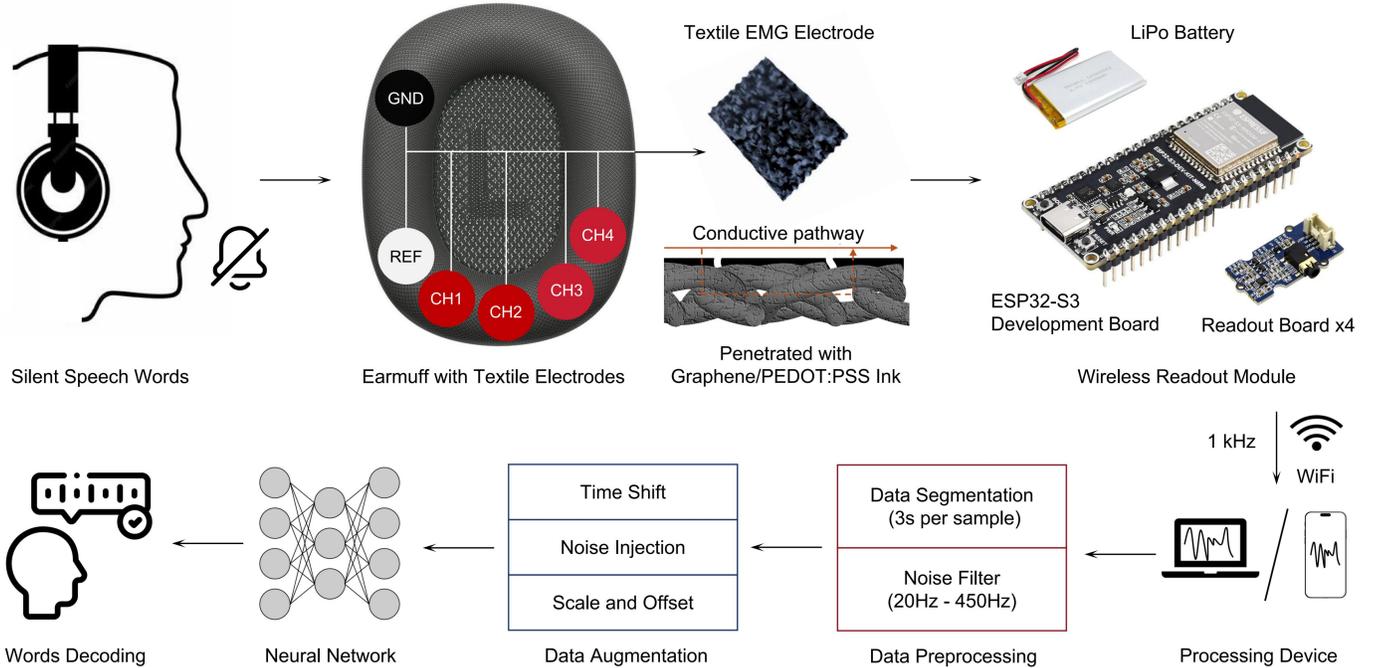

**Fig. 1. Diagram of the proposed headphone SSI system.** The system integrates multi-channel textile EMG electrodes into the earmuffs of commercial headphones, enabling real-time silent speech decoding. The electrodes, coated with graphene/PEDOT:PSS composite, provide high conductivity and breathability, enhancing skin-electrode coupling. A wireless readout module with four EMG readout circuits and an ESP32-S3 microcontroller processes signals at 1 kHz sampling rate and transmits data via Wi-Fi to a PC or mobile device. The software pipeline includes data preprocessing (segmentation, bandpass filtering), data augmentation (time shift, noise injection, scaling), and a 1D SE-ResNet classifier. The SE mechanism adaptively weights EMG channels, improving robustness against contact variations, enabling real-time silent speech decoding for human-machine interaction.

### 1) Textile Towel-Based EMG Electrodes

The textile-based EMG electrodes were fabricated using a graphene/PEDOT:PSS composite coating applied to high-porosity towel microfibers. As confirmed by scanning electron microscopy (SEM) analysis, the soak-coating process ensures uniform conductive material deposition, forming a conformal layer over individual microfibers while preserving the intrinsic flexibility and breathability of the textile substrate [22, 23]. Compared to conventional gel electrodes, towel-based electrodes provide superior air permeability and wearer comfort, making them ideal for discreet and prolonged use in daily scenarios.

The skin-electrode impedance characteristics were evaluated using a two-electrode setup with a potentiostat (EmStat4S, PalmSens). The results indicate that at low frequencies (1–10 Hz), the textile electrodes exhibit slightly higher capacitance, likely due to air gaps between the fibers and the skin. However, in the EMG-relevant frequency range (10–150 Hz), the impedance of the textile electrodes is comparable to that of wet electrodes, demonstrating effective charge transfer and stable biopotential recording. The high surface area and porous structure of the textile electrodes contribute to an enhanced effective contact area, thereby mitigating contact instability and motion artefacts.

### 2) Readout Module and Wireless Data Transmission

To enable real-time multi-channel EMG signal acquisition, the system employs four SeeedStudio EMG readout circuits, interfaced with an ESP32-S3 microcontroller unit (MCU). The entire readout module, including EMG amplifiers, analog-to-digital conversion (ADC), and wireless transmission components, is compactly integrated into the headphone earmuff housing, maintaining the original form factor of consumer headphones.

Power is supplied via a LiPo battery, ensuring untethered operation with minimal power consumption. The digitized EMG signals are wirelessly transmitted via Wi-Fi to an external computing device, such as a PC or smartphone, where real-time silent speech decoding is performed. The computational cost of the proposed 1D SE-ResNet model is optimized, requiring $O(10^7)$ floating-point operations (FLOPs) per inference, making it feasible for execution on mobile and edge devices without significant latency, i.e. in the order of µs for a CPU with a processing efficiency of 10 GFLOPs/ms, common in modern smartphones and PCs.

### B. Software of the System

#### 1) Preprocessing

The raw EMG signals collected from the four textile-based towel electrodes undergo a structured preprocessing pipeline to enhance signal quality before feature extraction and classification. The first step in this pipeline is data segmentation, where the continuous EMG recordings are divided into 3-second time windows, each corresponding to a single silent speech command. Given the 1000 Hz sampling rate, each segment consists of 3000 time points per channel, forming a multi-channel time-series dataset.

To mitigate noise and improve the signal-to-noise ratio (SNR), a 4th-order Butterworth bandpass filter (20–450 Hz) is applied to each channel. This frequency range is chosen based on the physiological properties of EMG signals, where most speech-related EMG activity lies between 20 Hz and 400 Hz, while higher frequencies (>450 Hz) are primarily thermal and electronic noise. As shown in Figure 2, the unfiltered raw signal (Figure 2a) exhibits large low-frequency fluctuations (<20 Hz), which are effectively suppressed after filtering (Figure 2b).

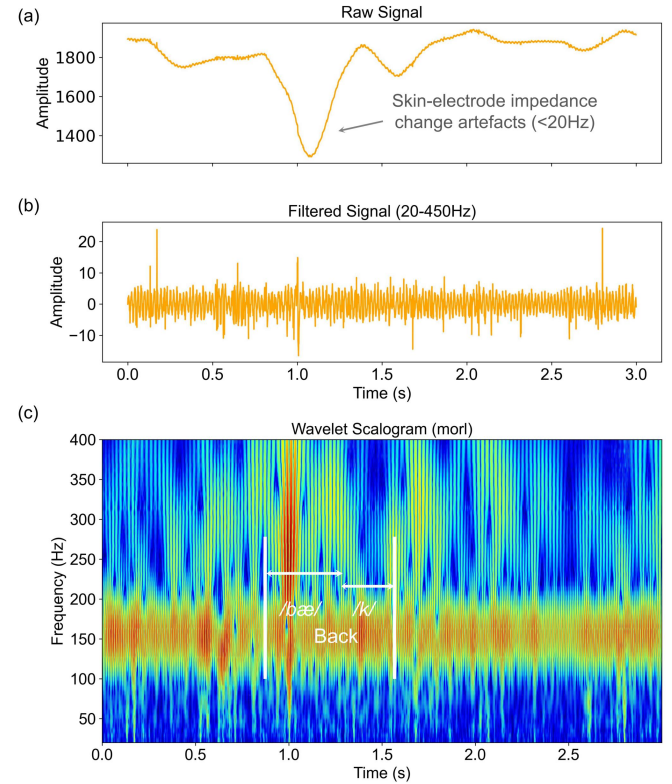

**Fig. 2. Preprocessing of the raw EMG signal.** (a) Example of the raw EMG signal recorded from a channel of the towel-based electrodes (with compact skin-electrode coupling), showing prominent low-frequency motion artefacts (<20 Hz) due to skin-electrode impedance variations. (b) Filtered EMG signal after applying a 4th-order Butterworth bandpass filter (20–450 Hz), effectively suppressing baseline drift and motion-induced noise while preserving neuromuscular activity. (c) Wavelet scalogram of the filtered signal, highlighting the preserved frequency components associated with silent speech articulation.

These low-frequency artefacts primarily stem from motion-induced impedance variations at the electrode-skin interface. Unlike traditional wet electrodes, which form a stable ionic conduction pathway, our textile-based dry electrodes offer enhanced comfort and breathability but introduce a tradeoff in signal stability, particularly due to (a) Variability in contact pressure - The earmuff-integrated textile electrodes do not have adhesive fixation, leading to contact fluctuations with head and jaw movements, which in turn cause time-dependent impedance changes; (b) Micro-slippage and triboelectric effects - The fibrous microstructure of the towel electrodes, while improving long-term wearability, allows for minor surface displacements, introducing low-frequency baseline drift; (c) Capacitive coupling instability - The presence of air gaps between the textile fibers and the skin affects charge transfer dynamics, introducing low-frequency capacitive artefacts that are highly variable.

Fortunately, these artefacts exhibit strong spectral separability, as their energy is concentrated predominantly below 20 Hz, whereas the informative EMG components reside in a much higher frequency range. This distinct separation allows for effective suppression of motion artefacts using a high-pass filtering approach, without significantly impacting the underlying neuromuscular signals relevant to silent speech decoding.

Figure 2c presents the wavelet scalogram of a filtered EMG signal, demonstrating the successful attenuation of low-frequency disturbances while preserving critical speech-related EMG activity. The artefact-reduced signal obtained after preprocessing serves as the input for subsequent feature extraction and classification, ensuring robust silent speech decoding.

### 2) Algorithm Development

To achieve robust and adaptive silent speech decoding, we develop a 1D SE-ResNet architecture, which integrates squeeze-and-excitation (SE) blocks within a residual network (ResNet) backbone to enhance channel-wise feature selection [24, 25]. As shown in Fig. 3, the model takes four-channel EMG signals as input and passes them through multiple convolutional layers, batch normalization (BN), and activation functions before classification into silent speech words.

The input to the model is a four-channel segmented EMG signal $X \in R^{4 \times T}$, where $T = 3000$ represents the number of time points per segment. The first layer applies a 1D convolution with kernel size $k$, stride $s$, and padding $p$, followed by a batch normalization and ReLU activation function:

$$H_1 = ReLU(BN(Conv1D(X, k, s, p)))$$

The core of our network is the SE-Res block, which consists of two sequential convolutional layers with identity shortcuts.

Given an intermediate feature representation $F \in R^{C \times T}$, the SE block first applies global average pooling (GAP) to squeeze temporal information into a channel descriptor:

$$z_c = \frac{1}{T} \sum_{t=1}^{T} F(c, t)$$

where $z_c$ is the aggregated feature for channel $c$. The SE block then learns adaptive recalibration weights using a two-layer fully connected (FC) network with a bottleneck ratio $r$:

$$s = \sigma(W_2 \cdot ReLU(W_1 z))$$

where $W_1 \in R^{C/r \times C}$ and $W_2 \in R^{C \times C/r}$, and $\sigma$ represents a sigmoid activation function. The final recalibrated feature map is obtained by element-wise multiplication:

$$F' = F \cdot s$$

This mechanism allows the model to assign higher importance to channels with better skin-electrode coupling, reducing the influence of noisy or poorly coupled channels caused by motion-induced impedance variations.

Following multiple SE-Res blocks, the final feature map undergoes global average pooling, a fully connected (FC) layer, and a softmax activation, producing a probability distribution over 10 silent speech words.

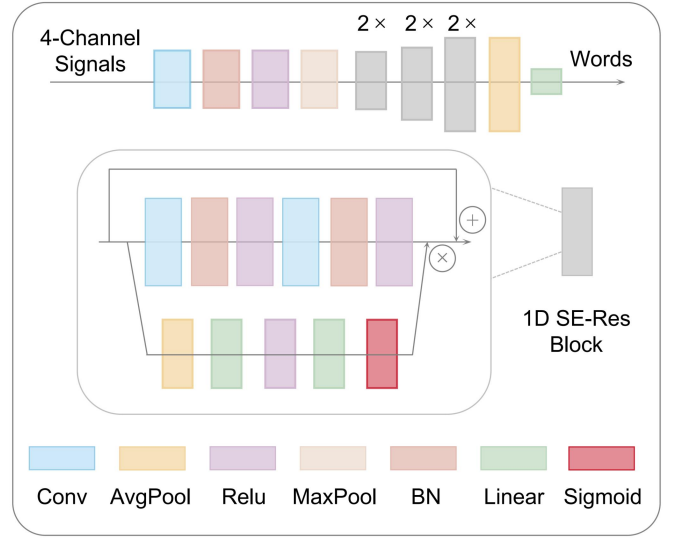

**Fig. 3. Diagram of the proposed 1D SE-ResNet.**

To enhance generalization and robustness, we apply three data augmentation strategies during training:

- **Time Shift**: Each EMG segment is randomly shifted along the time axis by up to ±100 ms, preserving sequence structure while introducing variability.

- **Noise Injection**: Gaussian noise with a signal-to-noise ratio (SNR) of 30 dB is added to simulate background electrical interference.

- Scale and Offset: Each sample is randomly scaled ($0.9\times$–$1.1\times$) and offset ($\pm0.1$ mV) to simulate variations in skin-electrode impedance.

These augmentation strategies introduce variability in signal dynamics, improving model robustness against real-world electrode coupling changes.

To ensure real-time feasibility, we optimize the computational complexity of the 1D SE-ResNet model. The complete inference pipeline, including preprocessing, feature extraction, and classification, operates within real-time constraints, making the system suitable for wearable applications. Pseudocode for the algorithm is provided below:

---

**Algorithm 1:** Ear-EMG Data Processing and Classification Pipeline (1D SE-ResNet)

**Input:** $\mathcal{D} = \{(X_i, y_i)\}_{i=1}^{N}$: dataset of $N$ samples, each $X_i \in \mathbb{R}^{4 \times 3000}$ (4 channels, 3 seconds, 1000Hz)
**Output:** Trained **1D SE-ResNet** model and predictions on new data

**Step 1: Preprocessing**
**foreach** *sample* $(X_i, y_i)$ **do**
  Apply a bandpass filter (20–450 Hz) on all 4 channels

**Step 2: Data Augmentation (Training Only)**
**foreach** *sample* $(X_i, y_i)$ *in training set* **do**
  Randomly apply:
- Time shift (up to 100 samples)
- Gaussian noise ($\sigma = 0.02$)
- Scaling ($0.9 \to 1.1$) and offset ($-0.1 \to 0.1$)

**Step 3: 1D SE-ResNet Model**
- Input: $\text{Conv1D}(4 \to 16, \text{kernel} = 7, \text{stride} = 2) \to \text{BatchNorm} \to \text{ReLU} \to \text{MaxPool}$
- Residual Blocks (3 stages):
  - Stage 1: $2\times \text{ResBlock}(16 \to 16, \text{stride} = 1)$
  - Stage 2: $2\times \text{ResBlock}(16 \to 32, \text{stride} = 2)$
  - Stage 3: $2\times \text{ResBlock}(32 \to 64, \text{stride} = 2)$
- Residual Block:
  - $\text{Conv1D} \to \text{BatchNorm} \to \text{ReLU} \to \text{Conv1D} \to \text{BatchNorm}$
  - **SE Block:** $\text{AdaptiveAvgPool} \to \text{FC(reduction} = 8) \to \text{ReLU} \to \text{FC} \to \text{Sigmoid} \to \text{Channel-wise scaling}$
  - $\text{Dropout} (p = 0.5) \to \text{Skip connection} \to \text{ReLU}$
- Output: $\text{GlobalAvgPool} \to \text{Dropout} (p = 0.5) \to \text{FC} (64 \to 10)$

**Step 4: Training Loop**
**for** epoch $= 1$ *to* maxEpochs **do**
  **foreach** *batch* $(X_b, Y_b)$ *from training set* **do**
    Compute $\hat{Y}_b = \text{1D-SE-ResNet}(X_b)$
    Compute cross-entropy loss $\mathcal{L}(\hat{Y}_b, Y_b)$
    Backpropagate and update weights
  Evaluate accuracy on validation set

**Step 5: Inference**
**foreach** *new sample* $X_{test}$ **do**
  Apply preprocessing and pass through model
  Predict class: $\arg\max(\hat{Y}_{test})$

---

## III. Results

### A. Collected Silent Speech Dataset

To evaluate the proposed headphone-integrated SSI system, we collected a dataset consisting of 10 commonly used control words. These words were selected based on their practicality in human-machine interaction (HMI) scenarios, including smart home control, assistive communication, and wearable device commands. Each word was repeated 100 times among the 4 subjects, forming a balanced dataset. 80% samples of the dataset was randomly split to the training set and 20% was the test set.

Table I presents the 10 words, their assigned IDs, and their International Phonetic Alphabet (IPA) transcriptions.

TABLE I
SILENT SPEECH WORD LIST AND PHONETIC TRANSCRIPTION

| ID | Word | IPA | ID | Word | IPA |
|---|---|---|---|---|---|
| 1 | Open | /ˈoʊpən/ | 6 | No | /noʊ/ |
| 2 | Close | /kloʊz/ | 7 | Next | /nɛkst/ |
| 3 | Start | /stɑːrt/ | 8 | Back | /bæk/ |
| 4 | Stop | /stɒp/ | 9 | Okay | /oʊˈkeɪ/ |
| 5 | Yes | /jɛs/ | 10 | Cancel | /ˈkænsəl/ |

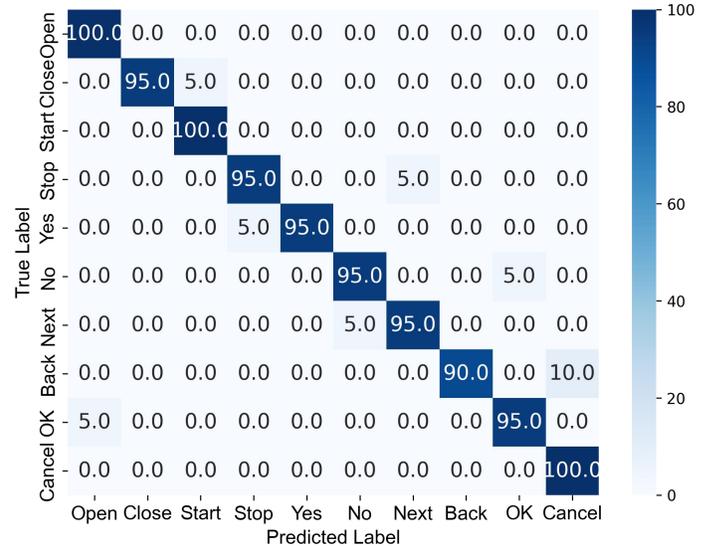

Fig. 4. Confusion matrix of the silent speech decoding.

### B. Model Performance

To assess the effectiveness of the proposed 1D SE-ResNet model, we trained and evaluated it on the collected silent speech dataset. The model achieved an overall classification accuracy of 96%, demonstrating its ability to robustly decode silent speech commands from four-channel EMG signals.

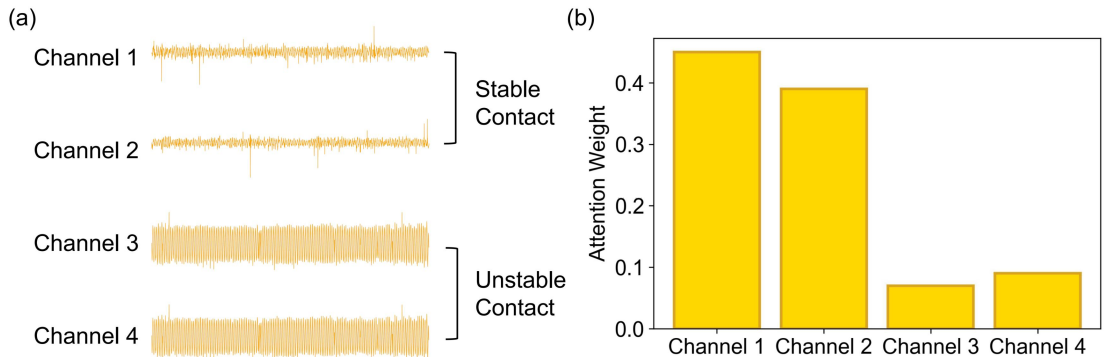

**Fig. 5. Attention-based adaptive channel selection in 1D SE-ResNet.** (a) Raw EMG signals from four channels, where Channel 1 and Channel 2 exhibit stable coupling, while Channel 3 and Channel 4 experience degraded signal quality due to weaker contact. (b) Extracted SE attention weights, confirming that the model assigns higher importance to well-coupled channels while suppressing noisy inputs, demonstrating effective adaptive channel selection.

Fig. 4 presents the confusion matrix of the classification results, illustrating the model's performance across different words. The matrix shows that most words were correctly classified, with some minor misclassification between phonetically similar words (e.g., "Back" and "Cancel"). Notably, words with distinctive phonetic structures such as "Open" and "Stop" achieved near-perfect classification accuracy.

### C. Attention-Based Channel Selection in Multi-Channel EMG

To further validate the adaptive channel selection capability of our 1D SE-ResNet, we conducted an explainable AI (XAI) analysis to examine how the model dynamically assigns importance to each EMG channel under varying skin-electrode coupling conditions. The attention values were extracted from the squeeze-and-excitation (SE) blocks, which compute channel-wise descriptors based on the overall feature activation of each channel. The SE mechanism first aggregates temporal information from each channel and then passes the aggregated features through a fully connected transformation with nonlinear activation functions, ultimately generating attention weights that scale the channel contributions. This enables the model to adaptively emphasize well-coupled channels while suppressing less reliable ones.

As shown in Fig. 5, the extracted EMG signals from four channels (Fig. 5a) demonstrate that Channel 1 and Channel 2 exhibit stable coupling, whereas Channel 3 and Channel 4 suffer from weaker skin-electrode contact, leading to noisier signals with increased fluctuation. The corresponding attention weight distribution (Fig. 5b) confirms that the model assigns significantly higher weights to Channel 1 and Channel 2, while Channel 3 and Channel 4 receive much lower attention values, reducing their influence on the final prediction.

This result provides direct evidence that the 1D SE-ResNet effectively adapts to channel variability, addressing one of the key challenges in wearable dry-electrode EMG systems. By dynamically weighting the most reliable channels, the model enhances robustness against motion-induced impedance changes, ensuring consistent classification accuracy even in real-world usage where electrode contact conditions fluctuate over time.

### D. Performance Analysis and Ablation Study

To comprehensively evaluate the impact of preprocessing, multi-channel signal integration, and the SE block, we conducted an extensive performance comparison across different model configurations and further analyzed the latent feature representations using t-SNE visualization [26].

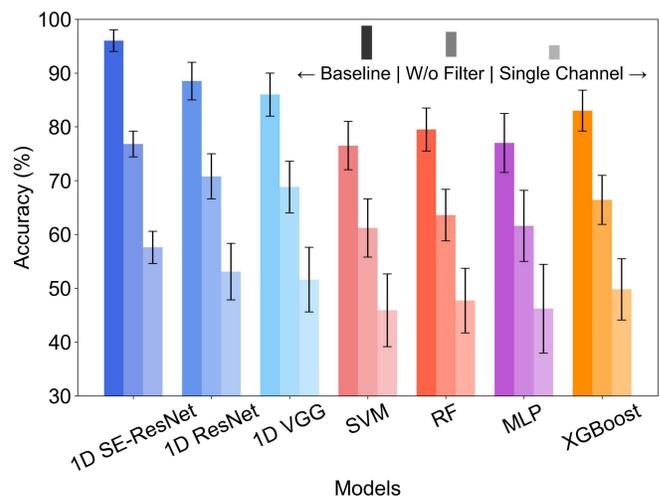

**Fig. 6. Model performance under different configurations.** The comparison highlights the necessity of motion artefact removal and multi-channel fusion for optimal classification accuracy. The proposed 1D SE-ResNet achieves the highest accuracy (96%) due to its ability to dynamically adapt to varying electrode-skin contact conditions.

#### 1) Model Performance Across Different Configurations

The classification results of 1D SE-ResNet and other benchmark models under three different settings are presented in Fig. 6. The Baseline configuration corresponds to the standard setup, which includes bandpass filtering (20-450 Hz) and four-channel EMG input. The W/o Filter configuration removes the bandpass filtering stage, leaving the raw signals unprocessed and thus retaining motion artefacts. The Single Channel configuration evaluates the performance when only the best-performing single-channel EMG input is used for classification.

Across all models, removing filtering significantly degrades classification accuracy, confirming the detrimental effect of motion artefacts on EMG decoding. The accuracy of 1D SE-ResNet drops from 96% to 76.5%, while traditional machine learning models such as support vector machine (SVM) and random forest (RF) suffer an even more substantial decline, highlighting their vulnerability to unfiltered noise. This demonstrates that motion artefacts introduce significant low-frequency distortions, which cannot be directly compensated by the classification model itself, further justifying the necessity of proper signal preprocessing.

Similarly, the Single Channel condition also leads to a considerable performance drop across all models. For instance, 1D SE-ResNet's accuracy declines by nearly 40% when using only a single channel, reinforcing the importance of multi-channel information fusion. In this case, models without channel selection mechanisms struggle to compensate for signal degradation due to unstable skin-electrode contact. The proposed SE mechanism in 1D SE-ResNet effectively mitigates this issue by dynamically adjusting attention weights to emphasize well-coupled channels while suppressing noisy or weakly coupled ones, ensuring consistent classification accuracy despite inevitable variations in electrode contact. These results confirm that both motion artefact removal and multi-channel integration play essential roles in achieving robust silent speech decoding in a headphone setting.

### 2) Model Performance Across Different Configurations

To further assess the feature extraction capability of 1D SE-ResNet, we conducted t-SNE visualization of the learned feature representations. Fig. 7 illustrates the distribution of data points in two different spaces: the raw EMG input space and the deep feature space extracted by our model. The raw input distribution exhibits significant overlap, indicating that the original EMG signals lack inherent separability, making direct classification challenging. In contrast, the deep features learned by 1D SE-ResNet exhibit well-formed clusters, demonstrating that the model successfully captures discriminative representations that enable precise classification of silent speech words.

The stark contrast between these two representations underscores the effectiveness of 1D SE-ResNet's hierarchical feature extraction mechanism. By leveraging deep convolutional layers and adaptive channel attention, the model transforms raw EMG signals into a structured latent space, where distinct speech categories are well separated. This confirms that the learned feature space provides strong inter-class separability, allowing the model to generalize effectively across different silent speech commands. These results further validate that deep learning-based feature extraction surpasses traditional handcrafted methods, reinforcing the need for advanced neural architectures in EMG-based silent speech interfaces.

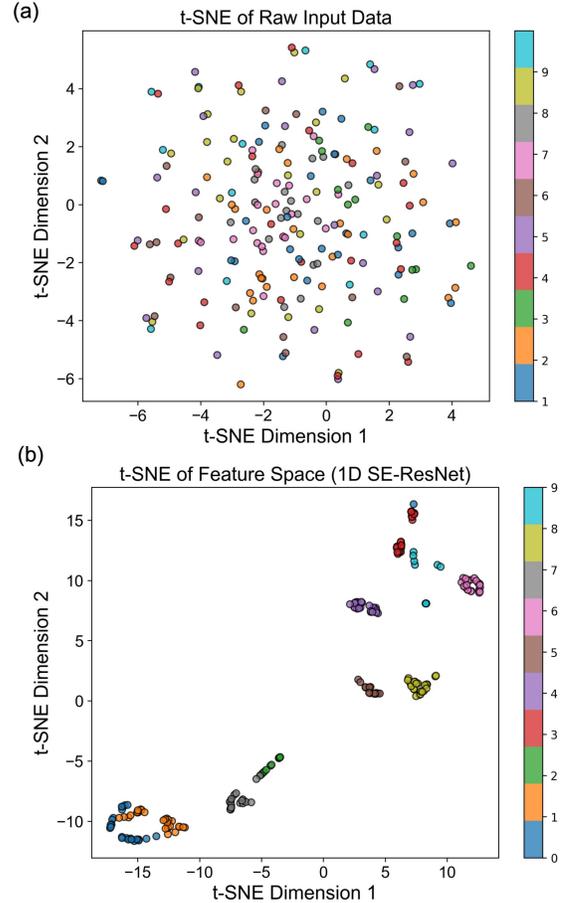

**Fig. 7. t-SNE visualization of feature representations.** (a) The raw EMG input lacks clear separability, leading to significant class overlap. (b) The extracted features from 1D SE-ResNet form distinct clusters, confirming its superior discriminative capability for silent speech classification.

### IV. DISCUSSION

The proposed headphone-integrated silent speech interface presents a novel approach that balances wearability, comfort, and usability, leveraging textile-based dry electrodes embedded into the earmuff. Unlike traditional patch-based EMG interfaces, which often require large electrode arrays applied directly to the face or neck, our approach offers a non-intrusive, user-friendly alternative while maintaining effective EMG signal acquisition. However, this integration also introduces inherent challenges related to electrode-skin coupling stability, as the contact pressure between the electrodes and the skin is more variable compared to adhesive-based wet electrodes. To address this limitation, we introduced a synergy of hardware and algorithmic enhancements, including towel-based textile electrodes with graphene/PEDOT:PSS coating to improve surface coupling, and a 1D SE-ResNet model that adaptively assigns higher attention weights to well-coupled channels while suppressing noisy signals. These advancements collectively enhance the system ' s robustness, ensuring reliable decoding of silent speech commands despite variations in contact quality.

Despite the promising results demonstrated in this work, several limitations remain, offering opportunities for future improvements. First, while we successfully integrated textile-based electrodes into the headphone earmuff, the current method relies on silver paste adhesion, which, while effective for prototyping, presents scalability challenges for mass production. Future research could explore direct printing, screen printing, aerosol jet printing, or laser-scribing techniques to fabricate seamless, high-resolution conductive electrode arrays directly onto the earmuff surface, enhancing manufacturability and durability [27, 28, 29, 30]. Second, the current vocabulary size and participant pool remain relatively limited; expanding the dataset to include a larger lexicon and a more diverse set of participants will be crucial for improving model generalization. Additionally, while our system demonstrates near-real-time inference capabilities, optimizing computational efficiency for on-device processing — such as implementing lightweight neural architectures for embedded platforms — would further enhance portability and real-world deployment [31, 32]. Finally, integrating context-aware learning could enhance robustness by allowing the system to dynamically adapt to user variability and environmental conditions. Combining silent speech decoding with large language models (LLMs) would enable intelligent interpretation of ambiguous commands, making interactions more fluid. Additionally, a multimodal approach incorporating eye-tracking, IMUs, and biosignals could further improve recognition accuracy, particularly in embodied AI and assistive robotics for applications like robotic teleoperation and exoskeleton control. Advancements in lightweight neural architectures and edge AI will be crucial for enabling real-time, on-device silent speech processing, paving the way for seamless human-machine interactions in wearable technology, accessibility, and hands-free communication [33, 34].

By addressing these aspects, future iterations of headphone-based EMG systems could move closer to seamless, real-world deployment, enabling intuitive silent speech communication in healthcare, assistive technologies, and hands-free control applications.

## V. Conclusion

This work introduces a novel headphone-integrated SSI that combines multi-channel textile-based EMG electrodes with an adaptive deep learning framework. By embedding dry towel electrodes into the headphone earmuff and leveraging the 1D SE-ResNet model for dynamic channel weighting, the system achieves 96% classification accuracy, ensuring robust and comfortable EMG signal acquisition. Its wireless and low-power design enhances practicality for real-world applications, enabling hands-free, voice-free interaction in assistive communication, human-computer interaction, and smart device control, paving the way for next-generation seamless human-machine interaction with the potential for integration into future multimodal and embodied AI systems.